\documentclass[]{aa}
\usepackage{graphicx}
\usepackage{txfonts}
\begin{document}
\title{AD Mensae: a dwarf nova in the period gap
\thanks{Based on observations collected at the European Southern Observatory, 
        La Silla, Chile and at the NOAO Cerro Tololo Interamerican Observatory, 
        Chile.}}
\titlerunning{AD Mensae: a dwarf nova in the period gap}
\author{L. Schmidtobreick \inst{1}
        \and
        C. Tappert \inst{2}}
\offprints{Linda Schmidtobreick, \email{lschmidt@eso.org}}
\institute{European Southern Observatory, Casilla 19001, Santiago 19, Chile.
           \and
           Departamento de Astronom\'{\i}a y Astrof\'{\i}sica, 
           Pontificia Universidad Cat\'olica, Casilla 306, Santiago 22, Chile}
\date{Received xxx xxx, xxx; accepted xxx xxx, xxx}
\abstract
{}
{AD\,Men has been classified as a probable long--period dwarf nova based on
its long--term variability. Recent spectroscopic data instead suggested a
short--period system. With the here presented observations we aim at
clarifying its nature.}
{Time--resolved photometry and spectroscopy has been 
used to get information on the orbital period of this system.}
{The light curve shows the typical flickering and a 
clear hump--like periodic modulation with an average amplitude of 0.3\,mag and
a period of $P=2.20(02)$\,h. The radial velocity measurements of the 
H$\alpha$ emission line confirm this value as the orbital period. }
{AD\,Men is thus located at the lower end of, but clearly inside, the gap 
of the period distribution of cataclysmic variables, making it one of only
11 dwarf novae in this important period range.}
\maketitle

\section{Introduction}
Dwarf novae (DNe) are a subtype of cataclysmic variables (CVs) and as such
are close, interacting binary systems, comprising a white dwarf receiving mass
from a Roche--lobe--filling late--type star. In DNe, mass transfer from the
secondary star to the primary takes place via an accretion disc, whose 
semi-regular long-term brightness variations are observed as ``outbursts''.
Several groups of DNe are distinguished according to their outburst behaviour,
the most prominent ones being the SS Cyg (or U Gem)- and SU UMa subtypes.
See Warner (\cite{warn95}) for a thorough introduction to these objects.

The evolutionary progress of a CV is reflected by the change of its orbital
period. Consequently, the period distribution of CVs represents the
observational testbed for theoretical models of CV evolution. One of the most
striking features of the observed distribution is a dearth of systems with
orbital periods between roughly 2 and 3 h, the so-called period gap. Generally,
this gap divides the systems with higher mass transfer rate, which are located 
at longer orbital periods, from the low mass transfer rate systems situated 
below the gap. 

AD\,Men has first been mentioned by Payne-Gaposchkin (\cite{payn71}) as a 
possible CV of SS\,Cyg subtype, i.e.\ as a dwarf nova with an orbital period
usually longer than 3 h. 

There was some confusion regarding
the identity of the object, as incorrect coordinates and finding charts 
were given in the old versions of the "Catalog and Atlas of 
Cataclysmic Variables" 
(Downes \& Shara \cite{down+93}; Downes et al. \cite{down+97}). 
Thus, also the wrong star was observed by Zwitter \& Munari \cite{zwit+95}. 
However, the online edition of this catalogue (Downes et al. \cite{down+01}) 
now contains the correct finding chart and coordinates.
Recently, spectroscopic data taken by Tappert \& 
Schmidtobreick (\cite{tapp+05}) confirmed the dwarf nova classification
for AD\,Men,
although several spectral characteristics suggested a period below the gap.

To clarify this apparent discrepancy, we have conducted time--resolved 
photometric and spectroscopic observations, in order to determine the orbital 
period of AD Men.

\section{The photometry}
\subsection{Observation and data reduction}
The photometric data were taken as part of a backup program in three nights on 
2004-12-27/29/30 using an R filter in front of a 512x512 CCD mounted at the
1.0 m SMARTS/CTIO telescope. The 
data reduction included the usual steps of bias
subtraction and division by skyflats. 

Aperture photometry for all stars on the CCD field was computed using the 
daophot package in IRAF and the stand\-alone daomatch and daomaster programs
(Stetson \cite{stet92}). Differential light curves were established with
respect to an average light curve including a total of 19 comparison stars
which were checked to be non--variable.

In this way we additionally discovered another, previously unknown, variable
star in the field at R.A. = 06:04:51.4 and DEC = $-$71:23:24 (2000.0) as 
measured from the Digitized Sky Survey (DSS). This object appears to be a
long--periodic variable, showing a smooth variation with an amplitude
within our observations of $\approx$ 0.6 mag.

Although our photometric data were not calibrated, a comparison with the
red DSS data, where AD Men clearly is caught during a bright state, shows
the target to be much fainter during our observations, i.e.\ at least close
to its recorded quiescence value.

\subsection{Results}
\begin{figure}
\resizebox{8.7cm}{!}{\includegraphics{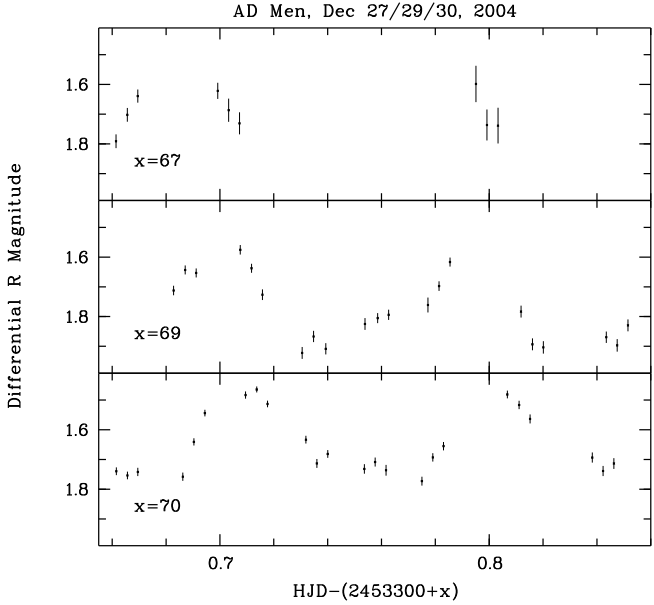}}
\caption{\label{lc}Differential R light curves from Dec 27, 29, and 30 (top to
bottom).}
\end{figure}

\begin{figure}
\rotatebox{-90}{\resizebox{!}{8.8cm}{\includegraphics{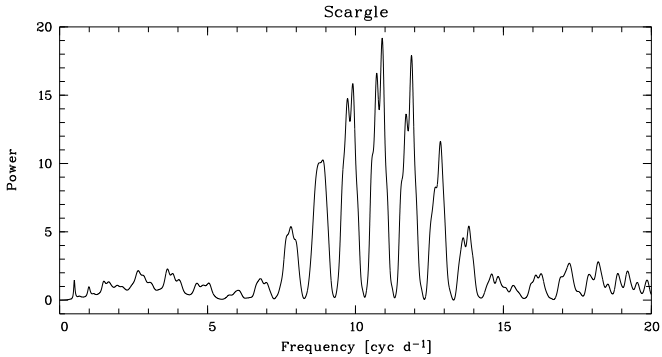}}}
\caption{\label{phpg}Scargle periodogram for the photometric data. The highest
peak corresponds to $f = 10.904~{\rm cyc~d}^{-1}$ and therefore $P = 0.0917$ d.}
\end{figure}

\begin{figure}
\rotatebox{-90}{\resizebox{!}{8.8cm}{\includegraphics{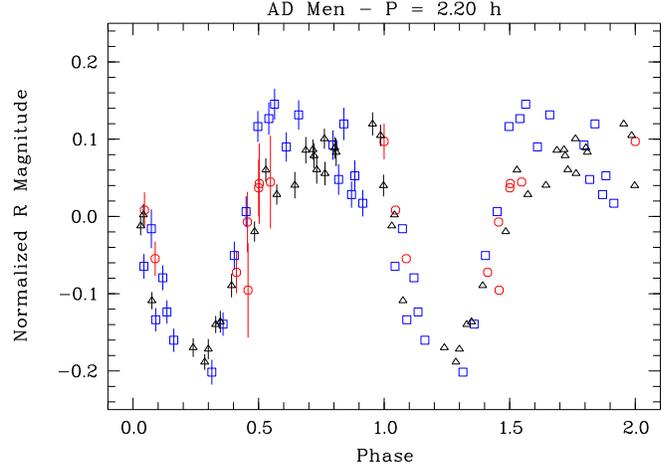}}}
\caption{\label{lc_phase}The differential magnitudes are plotted against their 
orbital phase with respect to the period $P=2.20$\,h, the first data point 
being arbitrarily set to phase zero. The individual symbols refer to the
different nights as follows: $\circ =$ Dec 27, $\Box =$ Dec 29,
$\bigtriangleup =$ Dec 30. Two phases are plotted for clarity, the second 
cycle without error bars.}
\end{figure}

The resulting light curve for AD Men is given in Fig.\ \ref{lc}. It clearly
shows the presence of a periodic feature in the form of a hump with an
amplitude of $\sim$0.3 mag. 
From this photometric variability, we can already confirm the 
assumption that AD\,Men is a system at rather high inclination 
(Tappert \& Schmidtobreick, \cite{tapp+05}). However a suspected inclination
as high as $80^\circ$ can be ruled out due to the absence of eclipses in the 
light curves.

In order to examine the periodicity, the
nightly average magnitude was subtracted from the individual light curves, and
the resulting normalised data set was analysed with the Scargle and
analysis-of-variance (AOV) algorithms implemented in MIDAS (Scargle 
\cite{scar82}, Schwarzenberg-Czerny \cite{schwa89}). Both methods agree
on exactly the same period $P_{\rm ph} = 0.0917(10)~{\rm d} = 2.20(02)~{\rm h}$
(Fig.\ \ref{phpg}). The uncertainties in these values were estimated by 
approximating the peak in the Scargle periodogram by a Gauss function and 
calculating $\sigma = {\rm FWHM}/2.355$. Fig.\ \ref{lc_phase} shows the 
normalised light curve folded with this period. While the width of the
hump is very similar for all three nights, its shape shows nightly variations,
probably due to flickering. 

There are two principal explanations for the periodic variation:

(1) The hump is caused by the bright spot, i.e.\ the impact site of the 
mass stream from the secondary star on the accretion disc. In this case,
$P_{\rm ph}$ represents the orbital period, placing the system inside the
period gap. Note that, although it is also possible that the periodic
variation 
is due to the
superhump phenomenon, we consider this
as very unlikely, since the system was at least close to quiescence
during our observations, and neither the spectroscopic data nor the
long-term behaviour indicate a high mass-transfer CV. 

(2) The light curve is the result of an ellipsoidal modulation due to a 
deformed, 
 elongated secondary, which gives a maximum in the light curve 
when seen along 
the minor axis and a minimum when seen along the major axis. 
In this picture, the orbital period would result as twice the 
photometric one: $P_{\rm orb}=4.4$\,h.

The presence of flickering somewhat favours the first 
possibility,
since the light from an accretion disc usually 
dominates
the contribution from the secondary star. 
Furthermore, following the method described by McClintock et al. 
(\cite{mccl+83}) with the limb-darkening values of Claret (\cite{clar98}),
we find that a very late--type secondary (M5 or later) would be needed, 
if the observed light variation with an amplitude of 
$\Delta m_R \approx 0.3$\,mag
was interpreted as being due to ellipsoidal variation.

Such a secondary is rather unlikely for a system with $P_{\rm orb}=4.4$\,h
where slightly earlier spectral types are expected
(Beuermann et al. \cite{beue+98}, Harrison et al. \cite{harr+05a}, 
and others herein).
Besides, a secondary star that dominates the orbital light curve should also
be easily visible in the optical spectrum. This, however, is clearly not 
the case (Tappert \& Schmidtobreick \cite{tapp+05}).

All evidence therefore points to the observed hump being caused by the
bright spot, and the orbital period resulting to $P = 2.2$\,h. 
However, since this places the system at an extraordinary position in the
period distribution of CVs, i.e.\ inside the sparsely populated
period gap, one would prefer a spectroscopic confirmation 
of this interpretation. 

\section{The spectroscopy}
We performed
follow--up spectroscopic observations with the aim of determining the
variation in radial velocities of the H$\alpha$ emission line. As these trace 
unambiguously the orbital motion of the emission--line source, 
they will confirm either of the two values for the orbital period.

\subsection{Observation and data reduction}
The observations were performed on 2005-01-27 using EMMI on the 
NTT at La Silla, Chile. A total of 54 spectra has been taken
with an individual exposure time of 300\,s. Thus, 
a time span of about 5\,h
has been covered, allowing to distinguish between the two 
possibilities
that are suggested by the light curves.

The standard reduction of the data has
been performed using IRAF. 
The overscan has been subtracted and the data have been divided by a flat field,
which was normalised by fitting Chebyshev functions of high order. 
The spectra have been optimally extracted (Horne \cite{horn86}).
Wavelength calibration yielded a final resolution of 0.39\,nm\ FWHM.

For all further analysis,
the MIDAS package and routines
written by the authors
have been used.

\begin{figure}
\rotatebox{-90}{\resizebox{!}{8.8cm}{\includegraphics{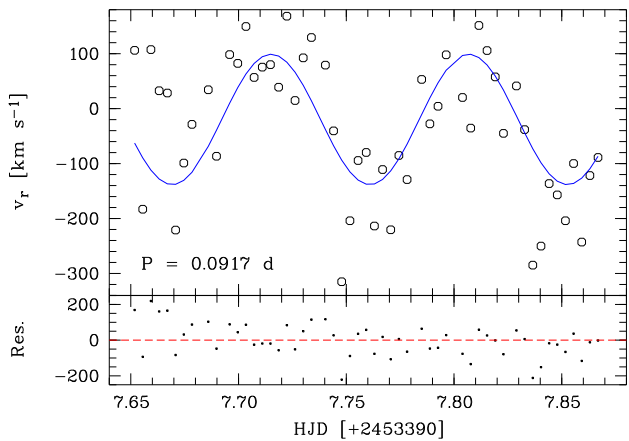}}}
\caption{\label{rv_plot}Upper plot: radial velocities of H$\alpha$ 
against heliocentric Julian date. The solid curve represents the best fit
using the photometric period. Lower plot: fit residuals.}
\end{figure}

\subsection{The radial velocity curve}

\begin{figure}
\rotatebox{-90}{\resizebox{!}{8.8cm}{\includegraphics{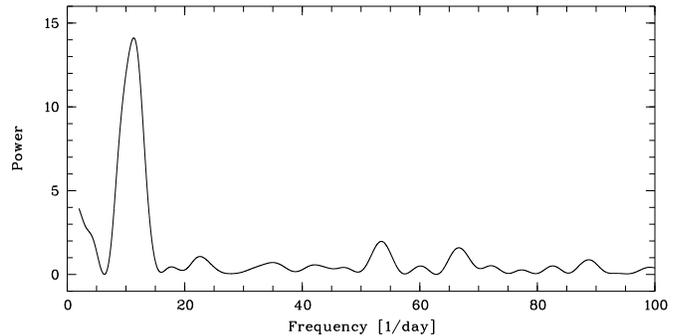}}}
\caption{\label{rv_scargle_plot}The power spectrum of the radial velocities
has been derived using the algorithm by Scargle (\cite{scar82}). The peak
value corresponds to an orbital period of $P=2.08(37)$\,h.}
\end{figure}



Since the original data mostly have very poor S/N due to variable weather
conditions, the spectra were smoothed by a continuous averaging over 7 pixels.
Radial velocities have then been determined by fitting a broad Gaussian
to the H$\alpha$ line. The resulting values are presented in 
Fig.\ \ref{rv_plot}.

Time-series analysis was performed using the Scargle and AOV algorithms 
implemented in MIDAS. As expected for data that cover one or 
two orbits within a single night, the periodograms show a broad maximum,
in both cases centred on a frequency corresponding to $P_{\rm sp} = 2.08(37)$ 
h. Within the errors, this value corresponds well with $P_{\rm ph} = 2.20(02)$ 
h, and we have therefore used this latter, more precise, value to fold the 
data. The best fit to the data was computed using a minimum-of-variance method 
on the sine function
\begin{equation}
v(\phi) = \gamma + K_1 \cdot \sin{\left( 2 \pi \left( \phi + 0.5 - \phi_0\right) \right) }
\end{equation}
where $v$ is the measured radial velocity and $\phi$ the phase with respect 
to $P_{\rm ph}$. The fitting parameters are the system velocity $\gamma$, the 
semi-amplitude $K_1$ of the radial velocity, and the phase shift $\phi_0$
for the red-to-blue crossing on the velocity curve with respect to the
first data point (HJD = 2\,453\,397.6518). 

The stability of the fit has been tested using
Monte-Carlo simulations which also yield the uncertainties of the
individual quantities.
We thus derive the following parameters: 
$\gamma = -20 \pm 8 $\,km/s, $K_1 = 119 \pm 10 $\,km/s,
and $\phi_0 = 0.94 \pm 0.01$.
Note that Tappert \& Schmidtobreick (\cite{tapp+05}) derived a lower limit
of $K_1 = 120$ km/s from spectra covering about 0.6\,h, in excellent agreement
with the here obtained value. 

The residuals corresponding to the fit show a slight linear slope over the 
observed time range (Fig.\ \ref{rv_plot}), which is probably due to the 
variation in airmass during the observations. Correcting for this slope 
yields $P_{\rm sp} = 2.12(39)$ h, and thus does not cause a significant change 
of the derived period within its 
accuracy.

\section{Conclusion}

\begin{figure}
\rotatebox{-90}{\resizebox{!}{8.8cm}{\includegraphics{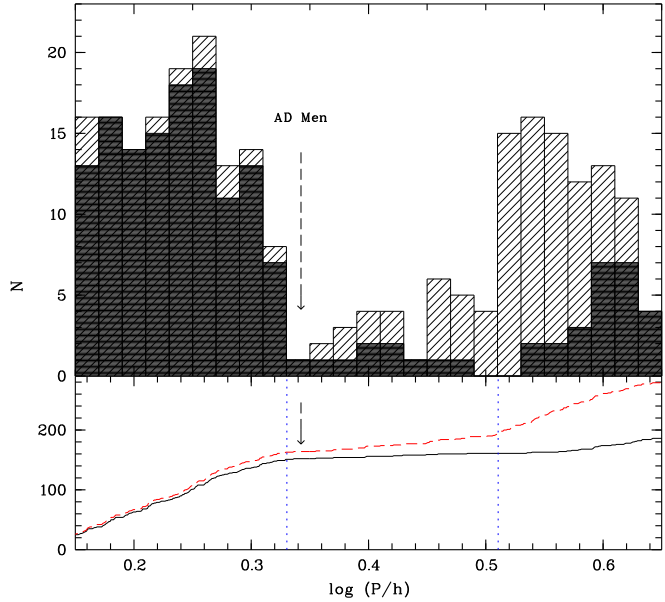}}}
\caption{\label{admen_gap} The period distribution of non--magnetic CVs is 
plotted between 1.4\,h and 4.5\,h. Hashed boxes indicate all non--magnetic CVs,
solid boxes indicate dwarf novae only.
The data were taken from the catalogue by Ritter \& Kolb (\cite{ritt+03}),
version 7.4. The position of AD\,Men, whose period is not 
included in either histogram, is marked by an arrow.
Below, the cumulative distribution is plotted for dwarf novae (solid line)
and all non--magnetic CVs (dashed line). The borders of the period
gap as defined in the text are indicated as vertical lines.
}
\end{figure}

\begin{table}[bt]
\caption{\label{gap_dn} AD Men and the known dwarf novae in the period gap 
according to the catalogue of Ritter \& Kolb (\cite{ritt+03}).}
\begin{tabular}{l l l l l}
\hline
 \noalign{\smallskip}
Object & Subtype & $P_{\rm orb}$ [d] & $P_{\rm orb}$ [h] \\
 \noalign{\smallskip}
\hline
 \noalign{\smallskip}
BX\,Pup     & Z Cam & 0.127    & 3.05 \\
V478\,Her   & SU UMa & 0.120 * & 2.88 \\
TU\,Men     & SU UMa & 0.117   & 2.81 \\
V405\,Vul   & SU UMa & 0.112 * & 2.69 \\
AX\,Cap     & SU UMa & 0.110 * & 2.64 \\
MN\,Dra     & ER UMa & 0.104   & 2.50 \\
J1702+3229  &        & 0.104   & 2.50 \\
NY\,Ser     & SU UMa & 0.098   & 2.35 \\
V725\,Aql   & SU UMa & 0.094 * & 2.26 \\
AD\,Men     &        & 0.092   & 2.20 \\
V589\,Her   & SU UMa & 0.091 * & 2.18 \\
 \noalign{\smallskip}
\hline
 \noalign{\smallskip}
\end{tabular}\\
{\footnotesize *: the orbital period has been  
estimated from the observed superhump period using the empirical relation 
given by Stolz \& Schoembs (\cite{stol+84}).}
\end{table}

The photometric and the spectroscopic data show variations that are
modulated with periods $P_{\rm ph} = 2.20(02)$ h and $P_{\rm sp} = 2.08(37)$ h,
respectively. The uncertainty in the spectroscopic period is such that it
still leaves some ambiguity with respect to the interpretation of the
photometric modulation. In principle, 
this modulation could be due to the occurrence of
superhumps from a precessing disc, and the orbital period would
then result a few percent shorter than $P_{\rm ph}$ (e.g., Patterson
\cite{patt98}). However, since the data do not provide further supporting
evidence for this assumption, we here take $P_{\rm ph}$ as the orbital
period.

Either case places AD\,Men at the lower edge of, but clearly within, the 
period gap of CVs (Fig.~\ref{admen_gap}). We here have defined the period gap 
from the Ritter \& Kolb (\cite{ritt+03}) data on the long period end via the 
general increase of CVs at an orbital period of 3.24\,h, and on the lower side 
at 2.14\,h. The gap appears broader for dwarf novae, reaching up 
to about 3.4\,h, with this upper limit not being well defined due to the low 
number of dwarf novae above the gap. AD Men now increases the number of
known dwarf novae in the period gap to 11 (Table\,\ref{gap_dn}), being one
of the six systems that have their orbital period measured directly. For the 
remaining five, this parameter has been derived from the observed superhump 
period, using the empirical relation given by Stolz \& Schoembs 
(\cite{stol+84}).

\section {Discussion}

The general assumption is that the systems in the period gap have formed there,
i.e.\ they have reached their CV configuration at an orbital period close
to the present one. 
Their secondary stars have not undergone the
structural changes that are supposed to lead to the cessation of mass transfer,
and thus to the existence of the period gap (Rappaport et al.\ \cite{rapp+83},
Spruit \& Ritter \cite{spruritt83}).

Instead, they probably have been acting as {\bf donors} for shorter times than
secondaries outside the gap.
Furthermore, they are supposed to be fully convective, thus a mixing between
the inner and outer layers of the star takes place.
Secondaries of CVs born above the period gap instead have undergone a 
stripping of the outer material, and {\bf expose} peculiar abundances as the 
inner layers become visible. Such abundances 
{\bf have been observed in non--magnetic CVs}
above the gap
(e.g. Harrison et al.\ \cite{harr+04a}; Harrison et al.\ \cite{harr+04b};
Harrison et al.\ \cite{harr+05a}).
For systems with an orbital period below 3\,h, such observations 
have been restricted to 
magnetic CVs (e.g. Harrison et al.\ \cite{harr+05a}), where indeed 
no peculiarities are observed.
However, this might be due to the magnetic nature of the stars
(see Harrison et al.\ \cite{harr+05b} for a discussion).

Detailed observations of the secondary stars in 
non--magnetic,
gap-born CVs are
still missing, basically due to the general difficulties in disentangling
this component from the other light sources in CVs
but there is hope that with the growing number of CVs in the gap, one
or the other will qualify for a more thorough research.


\acknowledgement{ 
We are grateful to Jorge Melnik for granting the time for the spectroscopic
follow--up observations 
and to Valentin Ivanov for conducting them in service mode. We acknowledge the
intense use of the SIMBAD database operated at CDS, Strasbourg, France.
This work has been partly supported by FONDECYT grant 1051078. We would also 
like to thank the referee, Hans Ritter, for valuable comments.
}

\end{document}